\documentclass[prd,aps,showpacs,preprintnumbers,amssymb]{revtex4}
\usepackage{graphicx}

\def\e3p{$\eta \rightarrow 3 \pi$}

\begin{document}

\title{%
\hfill{\normalsize\vbox{%
\hbox{\rm SU}
 }}\\
{Absolute neutrino masses}}

\author{Joseph Schechter $^{\it \bf a, b}$~\footnote[3]{Email:
 schechte@physics.syr.edu}}

\author{M. Naeem Shahid $^{\it \bf a}$~\footnote[4]{Email:
 mnshahid@physics.syr.edu}}

\affiliation{$^ {\bf \it a}$ Department of Physics,
 Syracuse University, Syracuse, NY 13244-1130, USA,}
 
 \affiliation{$^ {\bf \it b}$$CP{^3}$-Origins,
University of Southern Denmark, Campusvej 55, DK-5230 Odense M, Denmark}

\date{\today}

\begin{abstract}

We discuss the possibility of using experiments timing the propagation of neutrino beams 
over large distances to help determine the absolute masses of the three neutrinos.
 
\end{abstract}


\maketitle

\section{Introduction}
Determining the absolute values of the three neutrino masses is a 
subject of great interest.
Historically, Pauli  introduced the concept of 
the neutrino in order to explain why the energy of the emitted 
electron in beta decay did not have a single value as expected 
for a two body final state, but had a spectrum of energies (for different
observations of the process).
Fermi then suggested experimentally determining the mass of the neutrino 
by  precisely measuring the electron's energy spectrum in the endpoint 
region. This procedure has yielded 
 steadily decreasing upper bounds on neutrino mass for more than
 half a century. It has however
 turned out to be difficult to reach the range suggested by the cosmology 
 bound \cite{cosmobound} for the three different neutrinos now known to exist:
 
 \begin{equation}
 (m_1 + m_2 + m_3)c^2 \leq 0.3 eV
 \label{cosbound}
 \end{equation}
 
 This bound is based on the gravitational force exerted by 
 neutrinos in the universe rather than directly from neutrino kinematics..
  However, heroic efforts \cite{kartan1} - \cite{kartan3} are continuing to make steady progress on 
 finding the neutrino masses using the kinematics of the tritium beta decay reaction.

  One of course has the associated problem of finding the masses of all 
  three neutrino mass eigenstates.  
 The neutrino oscillation experiments \cite{superK} - \cite{minos} have determined
 these mass differences up to an ambiguity of either a ``normal" or an
 ``inverse" hierarchy. Specifically,
 
 \begin{eqnarray}
 A&=&(m_2)^2 -(m_1)^2 = (7.59\pm 0.21) \times 10^{-5} eV^2,
 \nonumber \\
  C&=& |(m_3)^2 - (m_1)^2| = (2.50\pm 0.13) \times 10^{-3} eV^2,
  \label{massdif}
  \end{eqnarray}
 
 where the numbers have been taken from the recent review \cite{RPP}.
 Of course, the two hierarchies correspond to the two possible signs of
 $(m_3)^2 - (m_1)^2$.
 Evidently, measuring any combination of three neutrino masses which is independent 
 of $A$ and $C$ should determine, for each hierarchy choice, the values of all three neutrino masses.
 
 \section{Toward finding the absolute neutrino masses}
 
  Here we would like to discuss some aspects related to the possibility of finding the     
  separate (or ``absolute") neutrino masses with the help of experiments 
measuring neutrino velocities by timing laboratory-made neutrino beams traveling 
from one point to another. At the moment, and for the near future, these are ``thought
 experiments". However, thinking about them, raises a number of interesting questions.

    If there were just a single neutrino it is clear that we could ``trivially" find its mass, $m$ 
    by measuring its velocity, $v$ as well as its energy, $E$ and using Einstein's formula:
    
   \begin{equation}
      E=\frac{m}{\sqrt{1-v^2}} \approx \frac{m}{\sqrt{2\epsilon}},
       \label{einfor}
       \end{equation}
       
        where we are working in units where the velocity of light, $c=1$ and $v=1-\epsilon$, appropriate to
         neutrinos traveling just slightly less than the speed of light.
   
    For definiteness, in the realistic 3-neutrino world, we assume a neutrino beam initiated
     in association with muons and detected, after traveling 
    a measured time and distance, by produced muons. We designate this as a mu-neutrino type beam.   
      Now we must take account of the fact that the laboratory neutrino beam in this setup corresponds
      to working with a neutrino ``flavor" 
   eigenstate (of muonic type) rather than a mass eigenstate. A prescription is therefore required to 
   define what might be called the ``averaged neutrino mass" $m$ which enters into Eq. (\ref{einfor})
   when we want to use it in the 3-flavor world.
   In Quantum Mechanics the mu-type neutrino corresponds to a  
   linear combination of the three mass eigenstates:
   
   \begin{equation}
        \nu_\mu = K_{21}\nu_1 +K_{22}\nu_2 + K_{23} \nu_3,
        \label{masseigens}
    \end{equation}    
         
where the $K_{2a}$'s are approximately the elements of the leptonic mixing matrix.
In models, say for the case of Majorana neutrinos, the mixing results from bringing
an``original" mass matrix $M$ to diagonal form $M_{diag}$,

\begin{equation}
M_{diag}=K^{T}MK,
\label{Mdiag}
\end{equation}

where $K$ is unitary.
The ``to-be-measured" mass, m will thus be interpreted \cite{ss1}  as the ``averaged"
quantity,

\begin{equation}
m = M_{22} = (K_{2i})^2 m_i,
\label{avmass}
\end{equation}

where the repeated index, $i$ should be summed over. We also have assumed for simplicity
that the $K_{ij}$'s are real. As a check of this formula, one notes that
if the neutrinos didn't mix we would have $K_{2i}=\delta_{2i}$ and $m = m_2$.
 
Present experimental information on the ``angles" which parameterize the matrix 
$K$ are summarized in ref. \cite{stv} 
Translating these angles to the needed matrix elements, $K_{2i}$ yields the numerical values with 
estimated errors:

\begin{eqnarray}  
K_{21}&=& 0.21 \pm 0.05
\nonumber \\
K_{22}&=& 0.68 \pm 0.05
\nonumber \\
K_{23}&=& 0.70 \pm 0.05 
\label{Knos}
\end{eqnarray}

Eq. (\ref{avmass}) then explicitly gives the to-be-measured quantity, $m$ 
as a different combination of the masses than those in Eq. (\ref{massdif}).
Thus combining the three equations we can find 
 the absolute neutrino masses for each hierarchy choice.
 For example, we may eliminate $m_2$ and $m_3$ in terms of $m_1$
 by using Eqs. (\ref{massdif}) to get 
 
 \begin{equation}
 m= (K_{21})^2 m_1 +(K_{22})^2\sqrt{A+(m_1)^2} +(K_{23})^2\sqrt{\pm C + (m_1)^2}.
 \label{m1eq}
 \end{equation}
 
 Here the plus sign corresponds to the normal hierarchy choice and the minus sign 
 corresponds to the inverted hierarchy choice. For the normal hierarchy we can, using this
 equation, obtain the curve in Fig 1 while for the inverted hierarchy we obtain the curve in Fig. 2.

\begin{figure}[htbp]
\centering
\rotatebox{0}
{\includegraphics[]{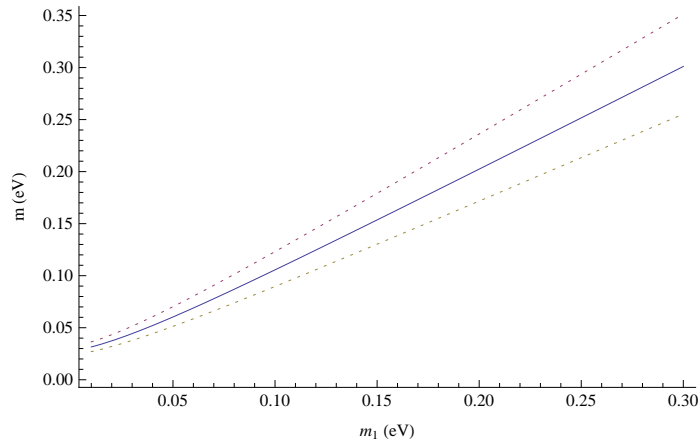}}
\caption[]{Plot of $m$ versus $m_1$ for normal hierarchy. The solid line corresponds to the mass $m$ ``measured" in a time of flight experiment. The two dotted lines give the error estimates for the crresponding  $m_1$.}
\label{mm1normal}
\end{figure}

  Thus if an average mass, m is measured by a long baseline time of flight experiment, we 
  can read $m_1$ from this curve and find also (from the known differences above), both $m_2$ and $m_3$.
  This should be done for both hierarchy assumptions; the present method does not determine which hierarchy 
  is the one nature chooses. 
  
     Of course, there are some important experimental errors introduced in this determination. It is 
     expected that further experimental measurements of neutrino oscillations will 
     substantially improve the experimental accuracy. 
    
    This procedure thus appears to be able to reasonably solve the problem of relating the ``averaged" mass, $m$
    measured by using Eq.(\ref{avmass}) to the three neutrino mass eigenvalues $m_i$.
     However, we are not done since we should pay some attention to the accuracy to which $m$ itself can 
    be determined.


    \begin{figure}[htbp]
\centering
\rotatebox{0}
{\includegraphics[]{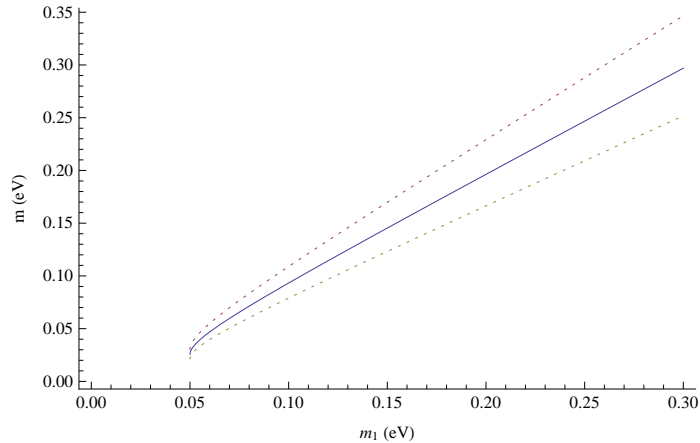}}
\caption[]{Plot of $m$ versus $m_1$ for inverted hierarchy. The solid line corresponds to the mass $m$ ``measured" in a time of flight experiment. The two dotted lines give the error estimates for the crresponding $m_1$.}
\label{mm1normal}
\end{figure}

    It should be remarked that the need to take mixtures of mass eigenstates into account
    is also evident in the electron endpoint of beta decay method for absolute neutrino mass determination.
     In that case the $K_{1i}$'s
   appear instead of the $K_{2i}$'s; see for example Eq. (21) of \cite{MNSTVW}.

    Next, we discuss the needed determination of the averaged mass, $m$ by a long baseline time of 
    flight measurement. First the mu-neutrino, $\nu_{\mu}$ must be produced. We would like to select the 
    $\nu_{\mu}$ to have as small an energy as possible in order that it travel at as low 
    a speed, $v$ as possible.
    
    An at least conceptually convenient source in this context might use the decay process:
    
    \begin{equation} 
     K^+\rightarrow \mu^+ + \nu_\mu.
     \end{equation}
 
    By selecting the $K^+$ and $\mu^+$ momenta one can reduce the energy of the $\nu_{\mu}$
    and hence its speed, $v$.
That would make a precise time measurement easier. From Eq. (\ref{einfor}), the deviation of the  neutrino's 
speed from that of light is given by:

\begin{equation}
\epsilon = 1-v = \frac{1}{2}(\frac{m}{E})^2.
\label{dev}
\end{equation}

If one were to select the neutrino energy to be 1 MeV, this would result in $\epsilon$ to 
be of the order of $10^{-14}$. That corresponds to a deviation from the velocity of light in 
the fourteenth decimal place, which does not seem practical.
 However, if one were able to get enough data with a neutrino 
energy selected to be 1 KeV, $\epsilon$ would be of the order of $10^{-8}$ which should be measurable.

\section{Summary}

We have shown how measurements of, for example, a mu-neutrino beam  velocity and energy can be used to help in
the determination of the absolute masses of the presently known three neutrinos. This could be a supplement to the 
determinations obtained using the endpoint of tritium beta decay method.

     Of course, an important question is the accuracy which can be achieved. In addition to the accuracy of the velocity 
     and energy measurements of the beam it is important to increase the accuracy of the neutrino mass differences and 
     the lepton mixing matrix elements obtained via neutrino oscillation experiments. A recent review of these is given in 
     \cite{stv}.

 \section*{Acknowledgments} \vskip -.5cm 

 This work was supported in part by the U. S. DOE under Contract no. DE-FG-02-85ER 40231. One of us (JS) would like to thank
  Prof. F. Sannino and the members of the CP$^3$ origins group at Southern Denmark University for their helpful discussions,
   warm hospitality and partial support..

\end{document}